\documentclass[twocolumn,showpacs,amsmath,amssymb]{revtex4}
\input{epsf}

\usepackage{graphicx}
\usepackage{array}
\usepackage{bigstrut}
\usepackage{longtable}
\usepackage{rotating,booktabs}
\usepackage{booktabs,threeparttable}
\usepackage{mathrsfs}
\usepackage{bm}

\begin{document}

\title{ Calculations of Bethe logarithm for Hydrogen and Helium using B-splines in different gauges}

\author{Yong-Hui Zhang$^{1}$, Lu-Jun Shen$^{1, 2}$, Chang-Min Xiao$^{2}$, Jun-Yi Zhang$^{1}$, and Ting-Yun Shi$^{1}$} 

\affiliation {$^1$State Key Laboratory of Magnetic Resonance and
Atomic and Molecular Physics, Wuhan Institute of Physics and
Mathematics, Chinese Academy of Sciences, Wuhan 430071, People's Republic of China}

\affiliation {$^{2}$Department of Physics, Hunan Normal University,
Changsha 410081, People's Republic of China}

\date{\today}

\begin{abstract}
The efficient and simple B-splines variational method is successfully used to calculate Bethe logarithm for the hydrogen atom in the velocity and length gauges. The ground state Bethe logarithm of hydrogen with fourteen accurate figures is obtained in the velocity gauge, and in the length gauge the ground state value has eleven accurate figures. Present velocity- and length-gauge results for the $ns$, $np$, $nd$, and $nf$ states up to $n = 200$ of hydrogen are at the 10$^{-10}$ level of accuracy, which represent the successful variational attempt to calculate Bethe logarithm of hydrogen in the velocity and length gauges. In addition, the B-splines variational method is successfully extended to calculate Bethe logarithm for the helium atom combined with configuration interaction. Results of numerical calculations for $n\,^{1,3}S$ up to $n=8$ states are presented in the acceleration gauge, the velocity gauge, and a hybrid of the velocity and acceleration gauges(also called pa-gauge). For the $2\,^3S$ state, the acceleration-gauge value of 4.364 036 7(2) a.u. and the pa-gauge value of 4.364 036 4(2) a.u. both have eight significant digits. For other triplet $S$ states, present results in three different gauges all have five to seven accurate figures. While for the singlet $S$ states, the best convergent values are obtained in the pa-gauge, of which the numerical precision is at the $10^{-5}$ to $10^{-7}$ level of accuracy as well.
\end{abstract}

 \maketitle

\section{introduction}

The experimental measurements and theoretical determinations of transition frequencies in atoms with one and more than one electron have advanced in accuracy to the point that they are sensitive to the QED contribution~\cite{parthey11, yost16, fleurbaey18, pachucki17}. At the leading term in the QED correction to the energy, one will encounter a quantity involving logarithm sum, which is called Bethe logarithm (BL from now on).

BL is one of the most complicated numerical evaluation quantities. Mainly because the logarithmic term $\ln|E_n-E_0|$ exists, where $E_0$ and $E_n$ are energies for the initial and intermediate states. Moreover there is no way to avoid the explicit dependence of BL on $(E_n-E_0)$ in one pure gauge. It is explained in detail that, on the one hand there will be a large contribution to BL from highly excited states, on the other hand, the negative contribution of the intermediate bound states will be canceled by contributions from continuum states with energies $E_n\geq1+E_0$, and this large cancelation requires an accurate representation of both the low- and high-energy regions of the spectrum~\cite{goldman84b,goldman94}, which certainly leads to the general variational techniques to collapse.

A variety of nonvariational methods~\cite{lieber68, huff69, drake90, jentschura05, jentschura05a} and variational methods of calculating the hydrogen atom BL in the acceleration or velocity gauge~\cite{goldman84b, haywood85, goldman94, mallampalli98, drake99, goldman00} have been proposed, and results with high accuracy have been achieved. In addition, the Gaussian basis set has been used in calculating the hydrogen atom BL as well~\cite{stanke13}, although giving a little poorer results than such as that obtained by the modified Slater-Laguerre radial functions~\cite{goldman00}. It must be noted that the spectral representation method in refs.~\cite{jentschura05, jentschura05a} are successfully finished in the length gauge, and abundant BL have been given for the ground state, low-lying and Rydberg states.

For the case of helium, in 1961 Schwartz first developed an integration representation to evaluate BL for the ground state of helium~\cite{schwartz61}, and his result has been the most accurate for more than 30 years~\cite{goldman83, goldman84b, bhatia98}. In 1999, Drake and Goldman suggested a direct variational method~\cite{drake99} to estimate BL for helium, in which the expression of BL was represented in terms of the acceleration gauge dipole operator. This method has been used to calculate BL for the ground and some low-lying states of helium~\cite{drake99, korobov04}, and the estimated accuracy is about nine to twelve significant digits. The same year, Korobov~\cite{korobov99} carried out the Schwartz-type expansion to compute BL for the $1\,^1S$ and $2\,^1S$ states of helium to a precision of 10$^{-6}$ to 10$^{-7}$. Korobov also has developed the Schwartz approach to deal with BL for a general three-body system such as helium~\cite{korobov12}, in which the dipole matrix elements are expressed in the velocity gauge. There have been other reports on the calculations of BL for the ground state and low-lying $S$ states of helium as well~\cite{baker93,bhatia98,baker00, yerokhin10, yerokhin18}.

B-splines~\cite{bachau01a, fischer08} have been successfully used to compute the hydrogen atom BL in the acceleration gauge~\cite{tang13}, which largely simplify the calculations and give high-accuracy results. This mainly dues to that B-splines can give good approximations to the bound and continuum states at the same time. As a follow-up to this work, in the present paper, we will calculate the hydrogen atom BL in the velocity and length gauges with the B-splines variational method. Moreover, we will extend B-splines variational method to compute BL for the helium atom combined with configuration interaction~\cite{chen92, chen93}, and calculations of BL for the $S$ states are carried out in the acceleration, velocity-acceleration and velocity gauges. Atomic units are used throughout this paper, and all calculations are finished in quadruple precision except special indications.

\section{formulations}
BL is defined as
\begin{eqnarray}
\beta=\dfrac{B}{C}
\,,\label{e1}
\end{eqnarray}
where
\begin{eqnarray}
B^{(A)}=\sum\limits_n\int|\langle\psi_0|\dfrac{Z\mathbf{r}}{r^3}|\psi_n\rangle|^2|(E_n-E_0)^{-1}\ln|E_n-E_0|
\,,\label{e2}
\end{eqnarray}
and
\begin{eqnarray}
C^{(A)}=\left \{
\begin{array}{cc}
\sum\limits_n|\int\langle\psi_0|\dfrac{Z\mathbf{r}}{r^3}|\psi_n\rangle|^2(E_n-E_0)^{-1}, &for\,\,\,s\,\,\,state,\\
\dfrac{2Z^4}{n_0^3},            &otherwise,\\
\end{array}
\right.
\label{e3}
\end{eqnarray}
are expressed in the dipole acceleration gauge. In Eqs. (\ref{e2}) and (\ref{e3}), $Z$ is the nuclear charge, $\mathbf{r}$ represents the position vector of the electron relative to the nucleus, the operator $\dfrac{Z\mathbf{r}}{r^3}$ will becomes $\sum\limits_i\dfrac{Z\mathbf{r}_i}{r_i^3}$ if there is more than one electron, $\psi_0$ and $E_0$ are the wavefunction and energy of the initial state with $n_0$ being the principle quantum number, and similarly $\psi_n$ and $E_n$ label the wavefunction and energy for one of a complete set of intermediate states. The summation integration over intermediate states includes the bound states as well as the continuum.

Based on the equivalent forms for the dipole transition matrix element~\cite{cowan81}, the expressions in the velocity and length gauges can be obtained respectively, where
\begin{eqnarray}
B^{(V)}=\sum\limits_n\int|\langle\psi_0|\mathbf{p}|\psi_n\rangle|^2|(E_n-E_0)\ln|E_n-E_0|
\,,\label{e4}
\end{eqnarray}
is in the velocity gauge with $\mathbf{p}$ being the momentum operator, and
\begin{eqnarray}
B^{(L)}=\sum\limits_n\int|\langle\psi_0|\mathbf{r}|\psi_n\rangle|^2|(E_n-E_0)^3\ln|E_n-E_0|
\,,\label{e5}
\end{eqnarray}
is in the length gauge. Expressions for $C^{(V)}$ and $C^{(L)}$ can be obtained according to Eqs.~(\ref{e4}) and (\ref{e5}) by removing the logarithm term $\ln|E_n-E_0|$. In addition, another expressions for BL in the other gauge which is called the pa-gauge are as following~\cite{goldman94, goldman00},
\begin{eqnarray}
B^{(VA)}&=&\sum\limits_n\int\langle\psi_0|\dfrac{Z\mathbf{r}}{r^3}|\psi_n\rangle
                          \langle\psi_n|\mathbf{p}|\psi_0\rangle
                          \ln\Bigg|\dfrac{\langle\psi_0|\dfrac{Z\mathbf{r}}{r^3}|\psi_n\rangle}
                                    {\langle\psi_n|\mathbf{p}|\psi_0\rangle}\Bigg| \nonumber \\
C^{(VA)}&=&\sum\limits_n\int\langle\psi_0|\dfrac{Z\mathbf{r}}{r^3}|\psi_n\rangle
                          \langle\psi_n|\mathbf{p}|\psi_0\rangle
\,.\label{e7}
\end{eqnarray}
The pa-gauge is a hybrid of the velocity and acceleration gauges, and can avoid the explicit inclusion of energies of the intermediate states. Different gauges can be used to monitor the calculations.

In order to calculate BL, we must obtain energies and wavefunctions firstly. In present calculations, the Hamiltonian of hydrogen is expressed as
\begin{eqnarray}
H=\dfrac{\mathbf{p}^2}{2}-\dfrac{Z}{r}\,,
\end{eqnarray}
and the hydrogen atom wavefunction is written as
\begin{eqnarray}
\psi_{n\ell m}(\mathbf{r})=|n\ell m\rangle=\sum\limits_i^NB_i^k(r)Y_{\ell m}(\hat{r})
\,,\label{e8}
\end{eqnarray}
where the radial component of the wavefunction is expanded with $N$ B-splines with the order of $k = 11$, and the angular component is a spherical harmonic function. The Hamiltonian of helium can be written as
\begin{eqnarray}
H=\dfrac{\mathbf{p}_1^2}{2}-\dfrac{Z}{r_1}+\dfrac{\mathbf{p}_2^2}{2}-\dfrac{Z}{r_2}+\dfrac{1}{r_{12}}\,,
\end{eqnarray}
and the helium atom wavefunction will be expanded with the following basis functions,
\begin{eqnarray}
\phi_{ij\ell_1\ell_2}(\mathbf{r}_1,\mathbf{r}_2,)=B_i^k(r_1)B_j^k(r_2)Y_{\ell_1\ell_2}^{LM}(\hat{r}_1,\hat{r}_2)\nonumber \\
         +(-1)^{\ell_1+\ell_2-L+S}B_j^k(r_1)B_i^k(r_2)Y_{\ell_2\ell_1}^{LM}(\hat{r}_1,\hat{r}_2)
\,,\label{e9}
\end{eqnarray}
where the radial basis functions are constructed in terms of B-splines of order $k = 7$, and $Y_{\ell_1\ell_2}^{LM}(\hat{r}_1,\hat{r}_2)$ is a coupled spherical harmonic which takes the form
\begin{widetext}
\begin{eqnarray}
Y_{\ell_1\ell_2}^{LM}(\hat{r}_1,\hat{r}_2)=\sum\limits_{m_1m_2}\langle\ell_1m_1\ell_2m_2|LM\rangle
        Y_{\ell_1m_1}(\hat{r}_1)Y_{\ell_2m_2}(\hat{r}_2)
\,. \label{e10}
\end{eqnarray}
\end{widetext}
The angular quantum numbers $\ell_1$ and $\ell_2$ in Eqs.~(\ref{e9}) and (\ref{e10}) will be chose as the number of partial wave $\ell_{max}$.

B-splines used to expand radial wavefunctions are defined by the exponential knots. The exponential knots $\{t_i\}$ with the exponential parameter $\gamma$ employed here will be the same with that used in ref.~\cite{zhang15},
\begin{widetext}
\begin{eqnarray}
\left \{
\begin{array}{ll}
t_i=0, &i=1,2,\cdot\cdot\cdot,k-1;\\
t_{i+k-1}=R_0\dfrac{\exp\left[\gamma R_0\left(\dfrac{i-1}{N-2}\right)\right]-1}{\exp[\gamma R_0]-1}, &i=1,2,\cdot\cdot\cdot,N-1;\\
&\\
t_i=R_0, &i=N+k-1,N+k\,.\\ %
\end{array}
\right.
\,\label{e11}
\end{eqnarray}
\end{widetext}
Here $R_0$ represents the radius of the box where the hydrogen or helium atom is placed in.
By diagonalization of the Hamiltonian matrix, we will obtain energies and wavefunctions for the initial and intermediate states to calculate BL.

\section{the hydrogen atom BL}
\begin{table*}[!htbp]
\caption{\label{t1} Results of BL (a.u.) for the ground state of hydrogen in the velocity and length gauges with the numbers of B-splines, $N$, and the exponential parameter of $\gamma$ changing.
The order of B-splines is $k = 11$, and the radius of the box is $R_0 = 50\,\,a.u.$ Calculations are finished in the double precision.}
\begin{ruledtabular}
\begin{tabular}{lllll}
 \multicolumn{1}{c}{$(N, \gamma)$}
&\multicolumn{1}{c}{$\beta^{(V)}(1s)$}                  &\multicolumn{1}{c}{$B^{(V)}(1s)$}
&\multicolumn{1}{c}{$C^{(V)}(1s)$}              &\multicolumn{1}{c}{$\delta C^{(V)}(1s)$}\\
\hline
(20,0.15)      &2.288         &4.576        &1.999 86          &1.38$\times10^{-4}$\\
(35,0.23)      &2.290 918     &4.581 83     &1.999 996 971 8   &3.03$\times10^{-6}$\\
(40,0.249)     &2.290 980 8   &4.581 963 66 &2.000 000 895 96  &8.96$\times10^{-7}$\\
(50,0.296)     &2.290 981 337 &4.581 962 83 &2.000 000 066 03  &6.60$\times10^{-8}$\\
\hline
\hline
 \multicolumn{1}{c}{$(N, \gamma)$}
&\multicolumn{1}{c}{$\beta^{(L)}(1s)$}          &\multicolumn{1}{c}{$B^{(L)}(1s)$}
&\multicolumn{1}{c}{$C^{(L)}(1s)$}              &\multicolumn{1}{c}{$\delta C^{(L)}(1s)$}\\
\hline
(20,0.09085)  &2.290 5       &4.59      &2.005     &5.08$\times10^{-3}$\\
(35,0.1316)   &2.290 93      &4.582 5   &2.000 29  &2.97$\times10^{-4}$\\
(40,0.14175)  &2.290 980 1   &4.582 3   &2.000 15  &1.53$\times10^{-4}$\\
(50,0.16348)  &2.290 981 43  &4.582 05  &2.000 039 &3.91$\times10^{-5}$\\
\end{tabular}
\end{ruledtabular}
\end{table*}
To begin with, we carry out a calculation of BL for the ground state of hydrogen as a test and demonstration case in the velocity and length gauges. Results of BL for the ground state of hydrogen of $\beta^{(V)}(1s)$ and $\beta^{(L)}(1s)$ , separately in the velocity and length gauges, with the numbers of B-splines, $N$, and the exponential parameter of $\gamma$ changing, are presented in Table~\ref{t1}, where the order of B-splines is $k = 11$ and the radius of the box is $R_0 = 50\,\,a.u.$ The values of $B^{(V)}(1s)$, $C^{(V)}(1s)$, $B^{(L)}(1s)$ and $C^{(L)}(1s)$ are also listed. $\delta C^{(V)(1s)} = |C^{(V)(1s)}-C_{exact}|$ and $\delta C^{(L)(1s)} = |C^{(L)(1s)}-C_{exact}|$, of which $C_{exact}$ is equal to 2. $C$ can be used to monitor the completeness of present B-splines basis. Calculations in Table~\ref{t1} are finished in the double precision.

From the top half of the table~\ref{t1}, it can be seen that with increasing $N$ and changing $\gamma$, the convergent rate of $C^{(V)}$ is somewhat better than that of $B^{(V)}$ because of the existence of the logarithm term, which limits largely the numerical precision of BL. The ground state BL of 2.290 981(1) with six convergent figures are obtained easily in present velocity-gauge calculations for $N = 50$ and $\gamma = 0.296$.

At the bottom half of the table~\ref{t1}, the length-gauge result of $\beta^{(L)}(1s)$ = 2.290 981(1) with six convergent figures is obtained. Because of the numerical cancellations, the convergence of $\beta^{(L)}$ is a little better than $B^{(L)}$. In addition, it indicates that the value of BL is very sensitive to the exponential parameter $\gamma$ for the relatively small dimension of B-splines basis in present length-gauge calculations of Table~\ref{t1}.

\begin{table*}[!htbp]
\caption{\label{t2} Comparisons of present value of BL (a.u.) with those given with the same B-splines fuctions, and those obtained using approximately the same number of basis functions. All values are compared to the exact group-theoretical result~\cite{huff69}. For the sake of comparison, only some selected figures given by Huff are shown in the last row. Present calculations are finished in the double precision.}
\begin{ruledtabular}
\begin{tabular}{llll}
 \multicolumn{1}{l}{Term}
&\multicolumn{1}{l}{N}        &\multicolumn{1}{l}{basis set}          &\multicolumn{1}{c}{$\beta(1s)$} \\
 \hline
$\beta^{(L)}$                           &50    &B-spline            &2.290 981 43    \\
$\beta^{(V)}$                           &50    &B-spline            &2.290 981 337   \\
$\beta^{(V)}$\cite{mallampalli98}       &102   &B-spline            &2.290 981 277   \\
$\beta^{(V)}$\cite{goldman00}           &54    &Slater-Laguerre     &2.290 945      \\
$\beta^{(V)}$\cite{stanke13}            &45    &Gaussian            &2.290 855 6864 \\
$\beta_{exact}$\cite{huff69}            &      &                    &2.290 981 375   \\
\end{tabular}
\end{ruledtabular}
\end{table*}

Table~\ref{t2} lists the comparison of present value of BL with those given with the same B-splines fuctions, and those obtained using approximately the same number of basis functions. All values are compared to the exact group-theoretical result~\cite{huff69}. For the sake of comparison, only some selected figures given by Huff are shown in the last row of Table~\ref{t2}. Using B-splines in the momentum space, ref.~\cite{mallampalli98} gave a velocity-gauge result of the same precision with our present value, but 102 B-splines are employed in their calculations. Goldman et al.~\cite{goldman00} achieved a value with five accurate figures by employing 54 modified Slater-Laguerre type basis, and Stanek et al.~\cite{stanke13} gave a result having four accurate figures using 45 Gaussian basis function. Present value obtained with 50 B-splines is more accurate than that of ref.~\cite{goldman00} by two orders of magnitude, and than that of ref.~\cite{stanke13} by three orders of magnitude.

\begin{table*}[!htbp]
\caption{\label{t3} Results of BL (a.u.) in the velocity and length gauges for the ground state of hydrogen using different exponential parameter, $\gamma$. $t_{12}$ represents the first non-zero inner knot, $E_{max}$ represents the energy
of the highest intermediate state.
The number of B-splines is $N = 300$ with the order of $k = 11$, the radius
of the box is $R_0 = 200\,\,a.u.$, and, a[b] represents $a\times10^b$. The digits in italics do not converge. Too many non-convergent digits of BL are listed in order to show the difference between values in two different gauges. }
\begin{ruledtabular}
\begin{tabular}{cllll}
 \multicolumn{1}{c}{$\gamma$}
&\multicolumn{1}{c}{$t_{12}$}                  &\multicolumn{1}{c}{$E_{max}$}
&\multicolumn{1}{c}{$\beta^{(V)}(1s)$}         &\multicolumn{1}{c}{$\beta^{(L)}(1s)$}\\
 \hline
0.005 &0.392    &4.44[3]  &2.25\emph{4 042 947}&2.25\emph{4 042 951}\\
0.015 &0.107    &5.96[4]  &2.280 \emph{895 920 439 8}&2.280 \emph{895 920 439 6}\\
0.026 &1.98[-2] &1.72[6]  &2.289 \emph{104 140 602 218 432 865}&2.289 \emph{104 140 602 218 432 849}\\
0.036 &3.73[-3] &4.81[7]  &2.290 6\emph{26 523 744 997 489 703}&2.290 6\emph{26 523 744 997489 704}\\
0.047 &5.45[-4] &2.23[9] &2.290 92\emph{9 303 425 834 593 785 2}&2.290 92\emph{9 303 425 834 593 785 8}\\
0.057 &9.04[-5]  &8.06[10] &2.290 972 \emph{712 409 531 547 2}&2.290 972 \emph{712 409 531 547 3}\\
0.067 &1.45[-5]  &3.09[12] &2.290 979 \emph{978 362 145 709 37}&2.290 979 \emph{978 362 145 709 31}\\
0.078 &1.89[-6]  &1.80[14] &2.290 981 1\emph{92 323 899 089 91}&2.290 981 1\emph{92 323 899 089 93}\\
0.088 &2.92[-7]  &7.52[15] &2.290 981 34\emph{6 903 679 587 95}&2.290 981 34\emph{6 903 679 587 92}\\
0.098 &4.45[-8]  &3.22[17] &2.290 981 370 \emph{882 427 71}&2.290 981 370 \emph{882 427 72}\\
0.109 &5.54[-9]  &2.06[19] &2.290 981 374 \emph{664 871}&2.290 981 374 \emph{664 872}\\
0.119 &8.27[-10] &9.16[20] &2.290 981 375 12\emph{4 593}&2.290 981 375 12\emph{4 592}\\
0.130 &1.01[-10] &6.06[22] &2.290 981 375 19\emph{5 6}&2.290 981 375 19\emph{5 8}\\
0.140 &1.49[-11] &2.78[24] &2.290 981 375 20\emph{4}&2.290 981 375 20\emph{2}\\
0.150 &2.18[-12] &1.28[26] &2.290 981 375 20\emph{5}&2.290 981 375 21\emph{8}\\
0.160 &3.18[-13] &5.99[27] &2.290 981 375 205 5\emph{21}&$-$\\
0.172 &3.14[-14] &6.10[29] &2.290 981 375 205 5\emph{50}&$-$\\
\hline
\hline
 \multicolumn{1}{c}{Term}
&\multicolumn{1}{c}{}  &\multicolumn{3}{l}{$\beta(1s)$}\\
 \hline
Present$^{(L)}$        &&\multicolumn{3}{l}{2.290 981 375 2\emph{18}}           \\
Present$^{(V)}$        &&\multicolumn{3}{l}{2.290 981 375 205 5\emph{50}}           \\
Ref.\cite{haywood85}   &&\multicolumn{3}{l}{2.290 981 375 205 6(1)}        \\
Ref.\cite{goldman00}   &&\multicolumn{3}{l}{2.290 981 375 205 552 301 342 5\emph{14}}\\
Ref.\cite{huff69}      &&\multicolumn{3}{l}{2.290 981 375 205 552 301 342 544 9686}   \\
\end{tabular}
\end{ruledtabular}
\end{table*}

It has been demonstrated the first nonzero inner knot can directly reflect the precision of BL in the B-splines variational method~\cite{tang13}. The ground state BL of hydrogen in the velocity and length gauges will be investigated with the first nonzero inner knot by changing the exponential parameter $\gamma$ at a large range. Here The box size is $R_0 = 200\,\,a.u.$, and the number of B-splines is $N = 300$ with the order of $k = 11$. Results are shown in Table~\ref{t3}, and the digits in italics do not converge.
In present calculations, the first nonzero inner knot is $t_{12}$, and $E_{max}$ represents the largest value in the intermediate energy spectrum. Too many non-convergent digits of BL are listed in order to show the difference between values in two different gauges.

Table~\ref{t3} shows that $\beta^{(V)}(1s)$ and $\beta^{(L)}(1s)$ have the same convergent results till $\gamma$ increased to 0.140, and achieve a value with ten convergent digits in both two gauges.  We can obtain a value of 2.290 981 375 205 550 a.u. with $\gamma$ equal to 0.172 in the velocity gauge. While for $\gamma > 0.150$, present quadruple precision program in the length gauge will give a divergent value because of the term of $(E_n-E_0)^3$ in Eq.~(\ref{e5}). This problem may be resolved to employ the much higher precision computational program.

The final convergent results are given in table~\ref{t3} as well, which are indicated by Present$^{(V)}$ and Present$^{(L)}$. Compared with the exact value given by Huff~\cite{huff69}, results with fourteen and eleven accurate figures are separately obtained in the velocity and length gauges, which are somewhat less accurate than the acceleration-gauge value given in ref.~\cite{goldman00}, but present velocity-gauge value is comparable to and even better than Haywood et al.'s~\cite{haywood85} obtained also in the velocity gauge. All calculations have provided an ample proof that high precision results of the hydrogen ground state BL can be obtained by B-splines variational method both in the velocity and length gauges.

\begin{figure}[!htbp]
\includegraphics[width=0.49\textwidth]{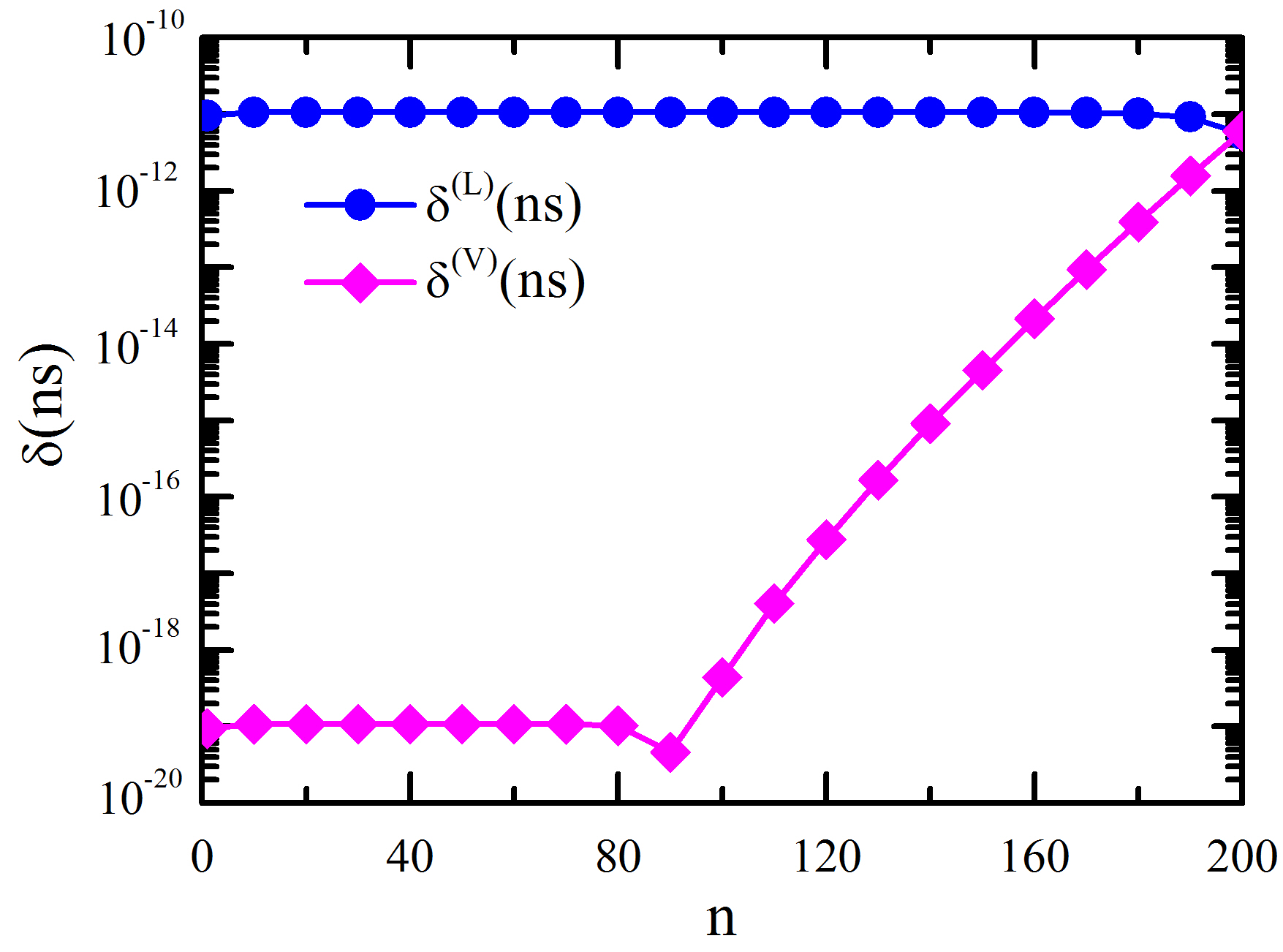}
\caption{\label{f1}(Color online) The differences between results of BL for ns(up to n=200) states of hydrogen in the velocity or length gauge and results in the acceleration gauge. $N = 3000$ B-splines with the order of $k = 11$ are used, the box radius is $R_0 = 110000\,\,a.u.$, and, the exponential parameter is $\gamma = 0.00027$. }
\end{figure}
\begin{figure}[!htbp]
\includegraphics[width=0.49\textwidth]{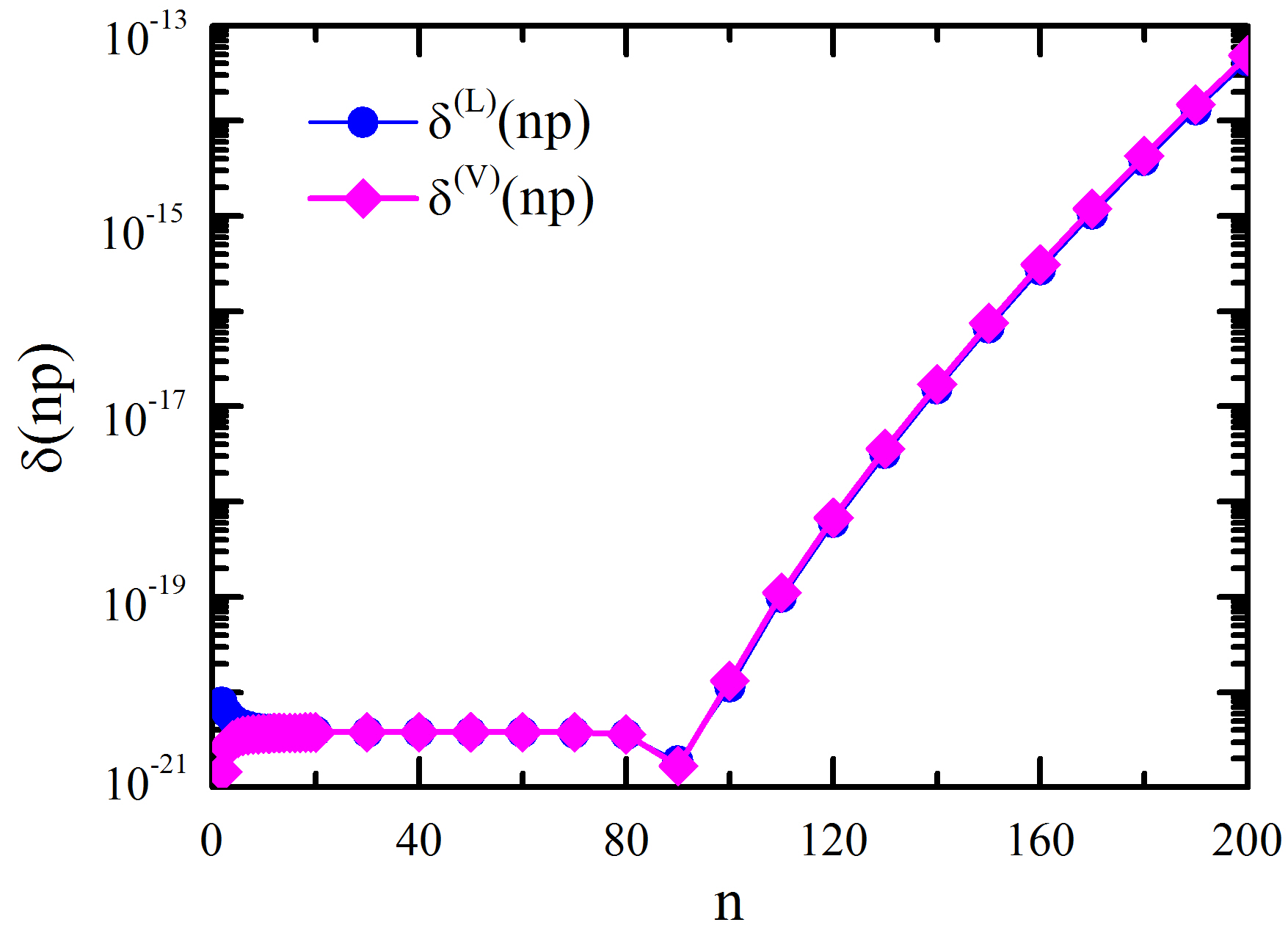}
\caption{\label{f2}(Color online) The differences between results of BL for np(up to n=200) states of hydrogen in the velocity or length gauge and results in the acceleration gauge. $N = 3000$ B-splines with the order of $k = 11$ are used, the box radius is $R_0 = 110000\,\,a.u.$, and, the exponential parameter is $\gamma = 0.0002$. }
\end{figure}
\begin{figure}[!htbp]
\includegraphics[width=0.49\textwidth]{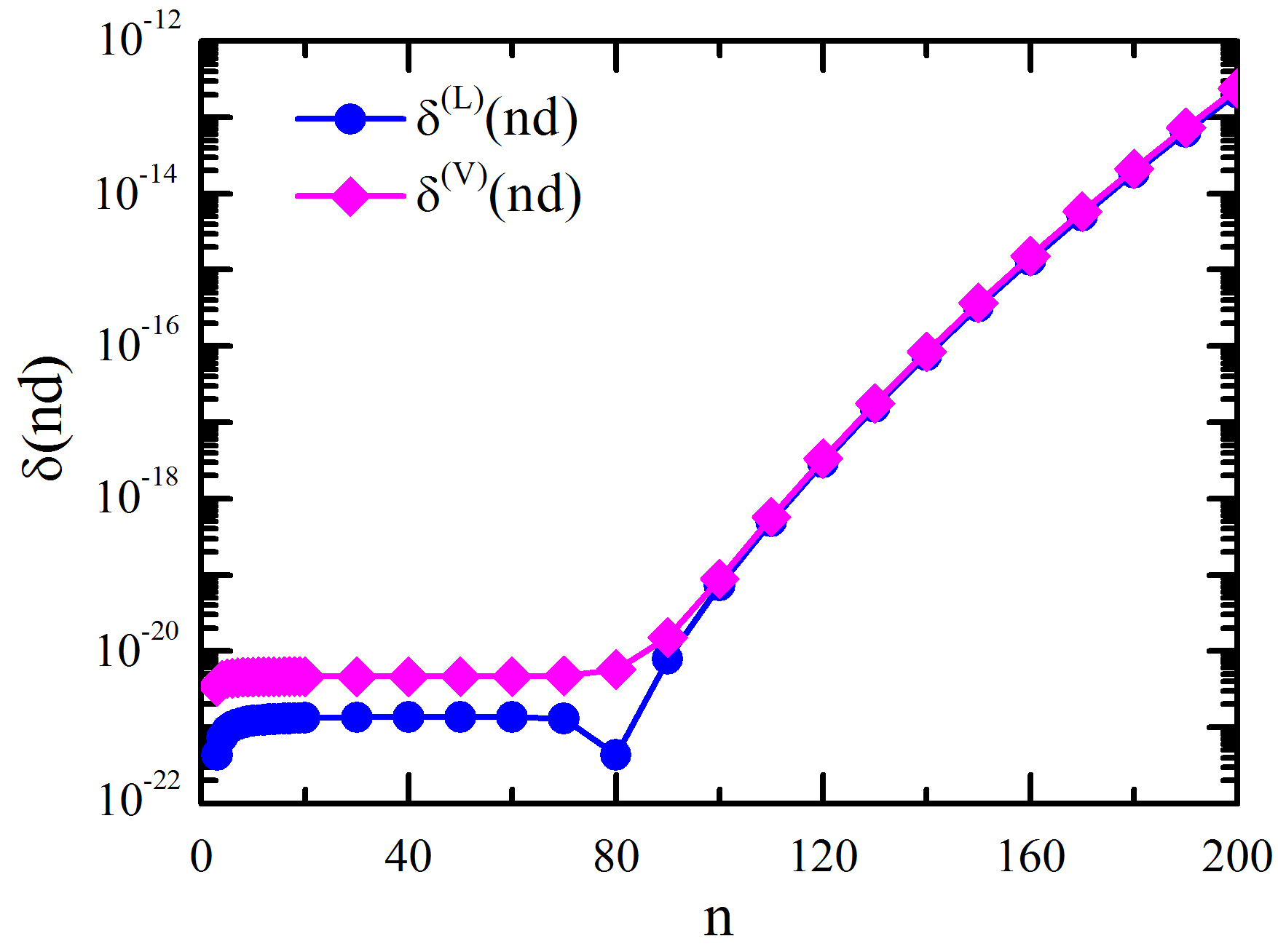}
\caption{\label{f3}(Color online) The differences between results of BL for nd(up to n=200) states of hydrogen atom in the velocity or length gauge and results in the acceleration gauge. $N = 3000$ B-splines with the order of $k = 11$ are used, the box radius is $R_0 = 110000\,\,a.u.$, and, the exponential parameter is $\gamma = 0.0002$. }
\end{figure}
\begin{figure}[!htbp]
\includegraphics[width=0.49\textwidth]{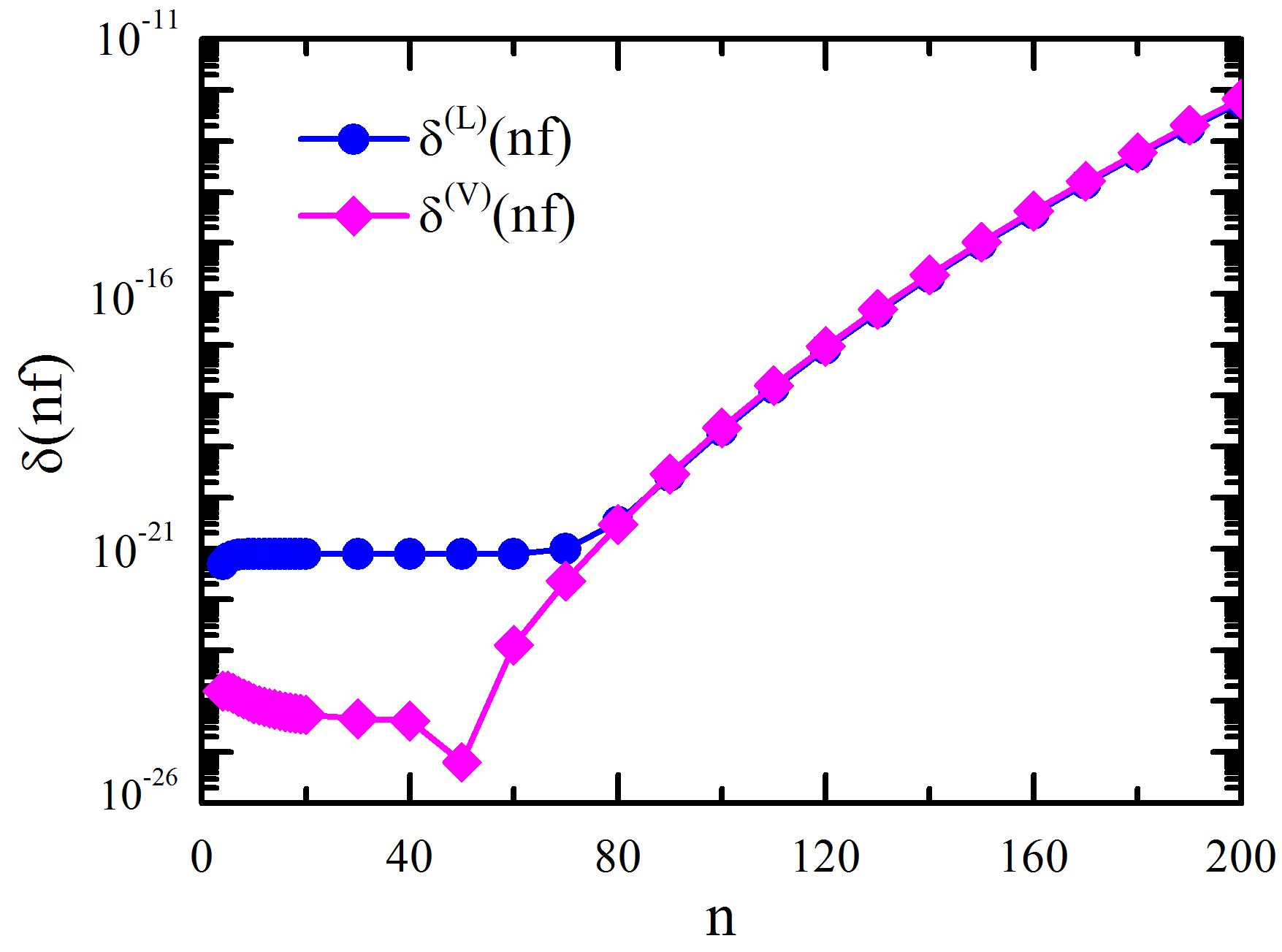}
\caption{\label{f4}(Color online) The differences between results of BLs for nf(up to n=200) states of hydrogen atom in the velocity or length gauge and results in the acceleration gauge. $N = 3000$ B-splines with the order of $k = 11$ are used, the confined box radius is $R_0 = 110000\,\,a.u.$, and, the exponential parameter is $\gamma = 0.0002$. }
\end{figure}

Then BL of $ns$, $np$, $nd$, and $nf$ states up to $n = 200$ are calculated based on the velocity- and length-gauge formulations in this subsection. The acceleration-gauge results are took as a benchmark. Quantities $\delta^{(V)}(n\ell)=\Big|\dfrac{\beta^{(V)}(n\ell)-\beta^{(A)}(n\ell)}{\beta^{(A)}(n\ell)}\Big|$ and $\delta^{(L)}(n\ell)=\Big|\dfrac{\beta^{(L)}(n\ell)-\beta^{(A)}(n\ell)}{\beta^{(A)}(n\ell)}\Big|$ are defined to analyze the agreement between different gauges. Results are graphically displayed in Figs.~\ref{f1}-\ref{f4}.
Present energy for the $200g$ state is $-$0.000 012 499 999 999 91 a.u. which approximates the exact energy $-1/2\times200^2 = -0.000 012 5$ a.u. very well. The first nonzero inner knot is at $10^{-10}$ orders of magnitude. It indicates that present values of BL should be accurate with about nine to ten figures. Present acceleration-gauge values have ten significant figures and can fully reproduce results shown in refs.~\cite{jentschura05, jentschura05a}. Figs.~\ref{f1}-\ref{f4} show that the agreement of values given in the velocity and acceleration gauges is better than those obtained in the length and acceleration gauges. But results under different gauges agree with each other better than $10^{-10}$ orders of magnitude, which implies that B-splines variational method has succeed in calculating BL for Rydberg states of Hydrogen in the velocity and length gauges, moreover has given high precision results with ten significant figures.

\section{the helium atom BL}

\begin{table*}[!htbp]
\caption{\label{t4} Convergence of BL (a.u.) for the $2\,^3S$ state of He in the acceleration gauge as the number of radial B-spline, $N$, and the angular partial wave, $\ell_{max}$, increased. B-splines with the order of $k = 7$ are used, the confined box radius is $R_0 = 20\,\,a.u.$, and, the exponential parameter is $\gamma = 0.855$. }
\begin{ruledtabular}
\begin{tabular}{cllllllll}
 \multicolumn{1}{c}{$N$}
&\multicolumn{1}{c}{$t_{8}$}
&\multicolumn{1}{c}{$\ell_{max}=1$}
&\multicolumn{1}{c}{$\ell_{max}=2$} &\multicolumn{1}{c}{$\ell_{max}=3$}
&\multicolumn{1}{c}{$\ell_{max}=4$} &\multicolumn{1}{c}{$\ell_{max}=5$}\\
\hline
25 &$1.59\times10^{-6}$&4.364 818 277 	&4.364 324 193  &4.364 298 525 	&4.364 295 000 	&4.364 294 244\\
30 &$1.09\times10^{-6}$&4.364 620 181 	&4.364 123 808 	&4.364 097 231 	&4.364 093 426 	&4.364 092 577\\
35 &$8.10\times10^{-7}$&4.364 570 208 	&4.364 073 039	&4.364 046 146 	&4.364 042 208 	&4.364 041 304\\
40 &$6.39\times10^{-7}$&4.364 567 781	&4.364 070 156 	&4.364 043 092 	&4.364 039 074 	&4.364 038 134\\
45 &$5.25\times10^{-7}$&4.364 566 734 	&4.364 069 037 	&4.364 041 912	&4.364 037 857 	&4.364 036 896\\
50 &$4.44\times10^{-7}$&4.364 566 669 	&4.364 068 937 	&4.364 041 785 	&4.364 037 711 	&4.364 036 737\\
Extrap.               &&&\multicolumn{4}{l}{4.364 036 7(2)}\\
Ref.~\cite{korobov04} &&&\multicolumn{4}{l}{4.364 036 820 3(1)}\\
Ref.~\cite{yerokhin10}&&&\multicolumn{4}{l}{4.364 036 820 41(3)}\\
\end{tabular}
\end{ruledtabular}
\end{table*}
For helium, we will take the $2\,^3S$ state as an example to examine the convergence of BL with the increasing number of B-splines, $N$, and the partial wave, $\ell_{max}$. Previous calculations suggest that good computational values of BL can be obtained with the magnitude of the first nonzero inner knot equal to $10^{-6}$ to $10^{-7}$~\cite{tang13}. In the present calculations, we will adjust the exponential parameter, $\gamma$, to make the first nonzero inner knot of $t_8$ at 10$^{-6}$ to $10^{-7}$ orders of magnitude. Table~\ref{t4} presents the convergence study of BL for the $2\,^3S$ state in the acceleration gauge. The helium atom is placed in a relative small box with $R_0 = 20$ a.u. because of only BL of the $2\,^3S$ state being considered. $\gamma$ is adjusted to be 0.855, and values of $t_8$ are listed in the second column of Table~\ref{t4}. As is seen from Table~\ref{t4} the convergence rate is somewhat higher as $N$ increased than as $\ell_{max}$ increased. This convergence style suggests that we can fix the partial wave of $\ell_{max}$, then increase the number of B-splines of $N$ to obtain our final convergent result. The extrapolated convergent value is given as 4.364 036 7(2) a.u., which has seven same figures with the best value given by Korobov and Yerokhin et. al.~\cite{korobov04, yerokhin10}. A conclusion can be drawn that B-splines have been used to calculate BL for the $2\,^3S$ state of He successfully, and also have given a result with high accuracy.

In the following, B-splines will be used to calculate BL for the low-lying excited $S$ states. Based on the convergence examination of Table~\ref{t4}, the number of the partial wave will be fixed as $\ell_{max} = 5$. In order to give a series of low-lying excited states from one diagonalization of Hamiltonian, a relative bigger box with the radius of $R_0 = 400$ a.u. will be chosen. We will firstly give the acceleration-gauge results. Calculations for the $n\,^{1,3}S$ states will be carried out in other two gauges as well, which one is the velocity gauge, and the other is a hybrid of the velocity and acceleration gauges. This pa-gauge avoids the explicit inclusion of the energies of the intermediate states, and consequently reduces the numerical round-off error in the $n$-th variational energy $E_n$~\cite{goldman94, goldman00}.
\begin{table*}[!htbp]
\caption{\label{t5} Results of BL (a.u.) for $n\,^3S\,\,(n = 2-8)$ states of He obtained in the acceleration, pa- and velocity gauges. The numbers in parentheses give the computational uncertainties. }
\begin{ruledtabular}
\begin{tabular}{clllll}
 \multicolumn{1}{c}{State}
&\multicolumn{1}{c}{$\beta^{(A)}$}
&\multicolumn{1}{c}{$\beta^{(VA)}$}
&\multicolumn{1}{c}{$\beta^{(V)}$}
&\multicolumn{1}{c}{Ref.~\cite{drake01}}
&\multicolumn{1}{c}{Ref.~\cite{drake99}}\\
 \hline
$2\,^3S$ &4.364 036 7(2)&4.364 036 4(2)&4.364 038(1)&4.364 035 417 &4.364 036 82(1) \\
$3\,^3S$ &4.368 666 7(1)&4.368 666 6(1)&4.368 667(2)&4.368 666 538 &4.368 666 92(2) \\
$4\,^3S$ &4.369 723(1)  &4.369 722 5(5)&4.369 723(1)&4.369 722 917 &4.369 723 441(5)\\
$5\,^3S$ &4.370 078(1)  &4.370 077(1)  &4.370 078(1)&4.370 079 109 &4.370 078 31(8) \\
$6\,^3S$ &4.370 228(4)  &4.370 227(1)  &4.370 227(2)&4.370 230 067 &                \\
$7\,^3S$ &4.370 30(2)   &4.370 300(6)  &4.370 298(7)&4.370 304 371 &                \\
$8\,^3S$ &4.370 33(5)   &4.370 33(2)   &4.370 32(3) &4.370 345 011 &                \\
\end{tabular}
\end{ruledtabular}
\end{table*}

Table~\ref{t5} lists the final convergent results of $n\,^3S\,\,(n = 2-8)$ states obtained in three different gauges, and the numbers in the parentheses are the computational uncertainties, which are given by the biggest difference between the extrapolated values and results obtained in the last three bigger basis sets.
For the $2\,^3S$ and $3^3S$ states, results in the acceleration gauge are one order of magnitude better than those in the velocity gauge. For other triplet states, the acceleration- and velocity-gauge values are at the same level of accuracy. The best convergence results in present B-splines calculations are obtained in the pa-gauge, which have five to seven accurate figures. This is mainly because that compared with wavefunctions, the uncertainties of BL are in large part from transition energies, and the pa-gauge calculations reduce the numerical round-off error in the $n$-th variational energy by avoiding the explicit inclusion of the energies.

We also give comparisons with Drake and Goldman's results~\cite{drake01,drake99} in Table~\ref{t5}, and the fifth column lists values obtained with the $1/n$ expansion of Eq.~(16) in ref.~\cite{drake01}, where BL for the $1s$ state of hydrogen is $\beta(1s) =$ 2.984 128 556. Compared with the correlated Hylleraas values given based on the acceleration gauge dipole operator~\cite{drake99}, present B-splines results in the acceleration and pa-gauges agree well with seven same figures for the $2\,^3S$ and $3^3S$ states. For the $4\,^3S$ and $5\,^3S$ states, present BL in the acceleration and pa-gauges have six same figures with the correlated Hylleraas values~\cite{drake99}. In addition, our ab-initio calculations of BL for the $6\,^3S$, $7\,^3S$, and $8\,^3S$ states in three different gauges all have five to six same figures with the $1/n$ expansion results~\cite{drake01}. It is concluded that for $n\,^3S\,\,(n = 2-8)$ states, the BL with five to seven accurate figures in three different gauges have been successfully achieved in present B-splines variational calculations.

\begin{table*}[!htbp]
\caption{\label{t6} Results of BL (a.u.) for $n\,^1S\,\,(n = 1-8)$ states of He obtained in the acceleration, pa- and velocity gauges. The numbers in parentheses give the computational uncertainties. }
\begin{ruledtabular}
\begin{tabular}{clllll}
 \multicolumn{1}{c}{State}
&\multicolumn{1}{c}{$\beta^{(A)}$}
&\multicolumn{1}{c}{$\beta^{(VA)}$}
&\multicolumn{1}{c}{$\beta^{(V)}$}
&\multicolumn{1}{c}{Ref.~\cite{drake01}}
&\multicolumn{1}{c}{Ref.~\cite{drake99}}\\
 \hline
$1\,^1S$ &4.370 34(2)  &4.370 14(2)    &4.370 6(4) &              &4.370 160 218(3)\\
$2\,^1S$ &4.366 43(1)  &4.366 412(1)   &4.366 5(2) &4.366 412 729 &4.366 412 72(7)\\
$3\,^1S$ &4.369 170(1) &4.369 164 3(2) &4.369 18(7)&4.369 164 888 &4.369 164 871(8)\\
$4\,^1S$ &4.369 893(1) &4.369 890 3(5) &4.369 90(2)&4.369 890 657 &4.369 890 66(1)\\
$5\,^1S$ &4.370 152(3) &4.370 151 1(2) &4.370 15(2)&4.370 152 093 &4.370 151 6(1)\\
$6\,^1S$ &4.370 27(1)  &4.370 266(2)   &4.370 26(2)&4.370 267 364 &\\
$7\,^1S$ &4.370 33(1)  &4.370 33(1)    &4.370 27(7)&4.370 325 649 &\\
$8\,^1S$ &4.370 34(4)  &4.370 34(2)    &4.370 34(2)&4.370 358 160 &\\
\end{tabular}
\end{ruledtabular}
\end{table*}
The finial convergent results obtained in three different gauges for the $^1S$ symmetry are given in Table~\ref{t6}. The BL except for the ground state listed in the fifth column are obtained using the $1/n$ expansion of Eq.~(16) in ref.~\cite{drake01}, and the correlated Hylleraas values~\cite{drake99} are listed in the last column. As is seen from the data of Table~\ref{t6}, the best convergence results for the singlet states are obtained in the pa-gauge, and the convergence style is better in the acceleration gauge than in the velocity gauge. Because of the more significant correlation effects, present results for the $1\,^1S$ and $2\,^1S$ states are less accurate than for the $3\,^1S$, $4\,^1S$ and $5\,^1S$ states. Compared with the previous results~\cite{drake01,drake99}, B-splines results except for the ground state agree within five to six significant digits in the acceleration gauge. In the pa-gauge, our best values have seven accurate figures with Drake and Goldman's values~\cite{drake99}. Present B-splines basis can not describe very well the electron correlation effects which are remarkable in the ground state, so the BL for the ground state are given as 4.370 34(2) a.u. and 4.370 14(2) a.u. respectively in the acceleration and pa-gauges, which only have four and five accurate figures. For the $6\,^1S$, $7\,^1S$ and $8\,^1S$ states, present calculations of BL have given direct and independent estimations, which are accordance with the $1/n$ expansion~\cite{drake01} at the $10^{-6}$ level of relative precision as well.

A solution to improve the ground state result is modifying the basis of Eq.~(\ref{e9}) with $r_{12}$ being included, similarly as done in the correlated Hylleraas basis. This termed Hylleraas-B-splines basis have been successfully used to calculate the static dipole polarizabilities of He~\cite{yang17}. And there is reason to believe that the ground state BL also can be given with good accuracy by using this modified B-splines basis~\cite{yang19}.

It is must be noted that for the helium atom, present B-splines variational calculation can not give correct convergent results of BL in the length gauge. This is mainly because we can not obtain energies with enough high accuracy, which only have five to seven accurate figures in present quadruple precision program. Compared with the hydrogen atom, our calculated energy for the $200g$ state is $-$0.000 012 499 999 999 91 a.u., so high-precision length-gauge values of BL for hydrogen still can be obtained. While present B-splines variational calculation of BL for helium indicates that the numerical precision of BL can be at the same level with energies of the corresponding states.

\section{conclusions}
In present paper, the hydrogenic ground and Rydberg states Bethe logarithm are calculated in the velocity and length gauges by using the B-splines variational method. We give a velocity-gauge result with fourteen accurate figures and a length-gauge value of eleven accurate figures for the $1s$ state of hydrogen. Bethe logarithm for Rydberg states in the velocity and length gauges are also achieved with high accuracy, which represents the successful variational attempt to calculate Bethe logarithm of the hydrogen atom in the velocity and length gauges.

In addition, B-splines variational method to calculate Bethe logarithm is successfully applied to helium combined with configuration interaction. Bethe logarithm for $^{1,3}S$ symmetry are calculated in three different gauges. For the triplet states, the best convergent values in three different gauges are at the $10^{-7}$ level of accuracy, while all other results given in present calculations also have at least five accurate figure. The best convergent results for the singlet states are achieved in the hybrid of the velocity and acceleration gauges, which have five to seven accurate figures as well.

\section{acknowlegments}
This work is supported by the National Natural Science Foundation of China under Grants No. 11704398 and No. 91536102, and by the Strategic Priority Research Program of the Chinese Academy of Sciences under Grant No. XDB21030300. Jun-Yi Zhang is grateful for the support from the ¡°Hundred Talents¡± program of the Chinese Academy of Sciences.


\end{document}